\documentclass[reprint,prd,twocolumn,superscriptaddress,nofootinbib]{revtex4-1}

\usepackage{color}
\usepackage{amsmath}
\usepackage{graphicx}
\usepackage{scalefnt}

\graphicspath{ {figs/} }

\allowdisplaybreaks

\begin{document}
\hfill TTP20-042,  P3H-20-075
\title{Third order corrections to the semi-leptonic \boldmath{$b\to c$} and the muon decays}
\author{Matteo Fael}
\email{matteo.fael@kit.edu}
\author{Kay Sch\"onwald}
\email{kay.schoenwald@kit.edu}
\author{Matthias Steinhauser}
\email{matthias.steinhauser@kit.edu}
\affiliation{Institut f\"ur Theoretische Teilchenphysik,
  Karlsruhe Institute of Technology (KIT), 76128 Karlsruhe, Germany}
%\date{}

\begin{abstract}
  We compute corrections of order $\alpha_s^3$ to the decay
  $b \to c \ell \bar\nu$ taking into account massive charm quarks. In the
  on-shell scheme large three-loop corrections are found. However, in the
  kinetic scheme the three-loop corrections are below 1\% and thus
  perturbation theory is well under control.  We furthermore provide results
  for the order $\alpha_s^3$ corrections to $b \to u \ell \bar\nu$ and the
  third-order QED corrections to the muon decay which will be important input for
  reducing the uncertainty of the Fermi coupling constant $G_F$.
\end{abstract}

\pacs{}
\maketitle

%- }}}
%- {{{ Intro.:

%\bigskip {\bf Introduction.}
\section{Introduction}
The Cabibbo-Kobayashi-Maskawa (CKM) matrix
determines the mixing strength in the quark sector and provides furthermore
the source for charge-parity (CP) violation in the Standard Model (SM).  It is
thus of prime importance to determine the parameters of the CKM matrix with
highest accuracy. In this article we address the elements $V_{ub}$ and
$V_{cb}$ which are accessible via semi-leptonic $B$ meson decays.

At present, the value of $|V_{cb}|$ from inclusive
$B\to X_c \ell \bar\nu$ decays is obtained from global
fits~\cite{Gambino:2013rza,Alberti:2014yda,Gambino:2016jkc}. The experimental
inputs are the semileptonic width and the moments of kinematical distributions
measured at Belle~\cite{Urquijo:2006wd,Schwanda:2006nf} and
BABAR~\cite{Aubert:2004td,Aubert:2009qda}, together with earlier data from
CDF~\cite{Acosta:2005qh}, CLEO~\cite{Csorna:2004kp} and
DELPHI~\cite{Abdallah:2005cx}. The most recent determination
in the so-called kinetic scheme
$|V_{cb}|=(42.19 \pm 0.78)\times 10^{-3}$~\cite{Amhis:2019ckw} has a relative
error of about 1.8\%, which is mostly dominated by theoretical uncertainties.
Global fits in the 1S scheme yield 
$|V_{cb}^\mathrm{1S}|=(41.98 \pm 0.45) \times 10^{-3}$~\cite{Bauer:2004ve,Amhis:2019ckw}.

A crucial ingredient for the determination of $|V_{ub}|$ and $|V_{cb}|$ is the
total semi-leptonic decay rate. Branching ratios of inclusive semileptonic $B$
mesons were measured at $B$ factories with a relative precision of about
2.5\%~\cite{TheBABAR:2016lja,Urquijo:2006wd,Mahmood:2004kq,Albrecht:1993pu}.
A relative uncertainty of 1.5\% is obtained with the help of a global fit:
$\mathrm{Br}(B \to X_c \ell^+ \nu_\ell) = (10.65 \pm
0.16)\%$~\cite{Amhis:2019ckw}.  Measurements are performed with a mild lower
cut on the electron energy~\cite{Urquijo:2006wd}, which excludes less than 5\%
of the events, or extrapolated to the whole phase space based on Monte
Carlo~\cite{TheBABAR:2016lja,Mahmood:2004kq}.  A key goal for Belle II is the
reduction of the systematic uncertainties on the branching fraction
determinations, as well as to obtain more precise and detailed measurements of
$B \to X_c \ell \bar \nu_\ell$ differential distributions~\cite{Kou:2018nap}.
Recent analyses by Belle and Belle II of leptonic and hadronic invariant mass
moments~\cite{q2moments,Abudinen:2020zwm} show that a percent or even
sub-percent relative accuracy can be achieved for certain observables.

With the help of the heavy quark expansion it can be written as a double
series in $\alpha_s$ and $\Lambda_{\rm QCD}/m_b$.  The $m_b$-suppressed
corrections are obtained from higher-dimensional operators. In the free-quark
approximation, corrections up to ${\cal O}(\alpha_s^2)$ are
available~\cite{Luke:1994yc,Trott:2004xc,Aquila:2005hq,Pak:2008qt,Pak:2008cp,Melnikov:2008qs,Biswas:2009rb,Gambino:2011cq,Dowling:2008mc}
together with the leading $\beta_0$ terms at higher orders~\cite{Ball:1995wa},
where $\beta_0$ is the one-loop coefficient of the QCD beta function.  The
power corrections of order $\Lambda_\mathrm{QCD}^2/m_b^2$ and
$\Lambda_\mathrm{QCD}^3/m_b^3$ have been computed
in~\cite{Chay:1990da,Bigi:1993fe,Manohar:1993qn,Gremm:1996df} to tree-level
and in~\cite{Becher:2007tk,Alberti:2013kxa,Mannel:2014xza,Mannel:2019qel} to
${\cal O}(\alpha_s)$.  Also $1/m_b^4$ and $1/m_b^5$ terms are known, however,
only at leading
order~\cite{Dassinger:2006md,Mannel:2010wj,Mannel:2018mqv,Fael:2018vsp}. Note
that linear $1/m_b$ corrections vanish to all orders.  Missing
higher-order perturbative and power corrections limit the current extraction
of $|V_{cb}|$.

The relative size of the second order corrections to the partonic
$b\to c \ell \bar \nu_\ell$ decays is about $1-3\%$ depending on the quark
mass scheme, with a theoretical uncertainty due to renormalization scale
variation estimated to be 1\%~\cite{Gambino:2011cq}, which soon can become
comparable to experimental errors.  In this work we make a major improvement
in the theory underlying $B\to X_c\ell\bar\nu$ decays by computing the
$\alpha_s^3$ corrections to the total rate, at leading order in $1/m_b$.  We
incorporate a finite charm quark mass via an expansion in the mass difference
$m_b-m_c$ and show that precise results can be obtained for the physical
values of $m_c$ and $m_b$.  Our analysis even allows for the limit $m_c\to0$
which provides $\alpha_s^3$ corrections for the decay rate
$\Gamma(B\to X_u\ell\bar\nu)$.\footnote{Note that in our approach one class of
  diagrams for the $b\to u$ transition is missing, namely the one where the
  charm quark appears as virtual particle in a closed loop. At
  ${\cal O}(\alpha_s^2)$ these corrections were denoted by
  $U_C$~\cite{Pak:2008qt,Pak:2008cp}.}

A process closely related to $b\to u\ell\bar\nu$ is the muon decay. Its
lifetime, $\tau_\mu$, can be written in the following form
\begin{eqnarray}
  \frac{1}{\tau_\mu} 
  &\equiv& \Gamma(\mu^-\to e^- \nu_\mu \bar{\nu}_e) 
  = \frac{G_F^2 m_\mu^5}{192 \pi^3}\left(1 + \Delta q \right)
      \,,
      \label{eq::tau_mu}
\end{eqnarray}
where $G_F$ is the Fermi constant, $m_\mu$ is the muon mass and $\Delta q$
contains QED and hadronic vacuum polarization corrections (see
Ref.~\cite{Kinoshita:1958ru,vanRitbergen:1999fi,Ferroglia:1999tg} for
details). Note that all weak corrections are absorbed in $G_F$.
Equation~(\ref{eq::tau_mu}) allows for the determination of $G_F$ if precise
measurements of $\tau_\mu$ are combined with accurate QED predictions.  We
compute for the first time $\alpha^3$ corrections to $\Delta q$ by specifying
the colour factors of our $b\to c\ell\bar\nu$ result to QED and taking the
limit $m_c\to0$.  This allows for the determination of the third-order
coefficient with an accuracy of 15\%.

%- }}}
%- {{{ Calculation:

%\bigskip {\bf Calculation.}
\section{Calculation}
\begin{figure}[t]
  \begin{tabular}{cc}
  \begin{tabular}{cc}
    \raisebox{2.2em}{\includegraphics[width=0.18\textwidth]{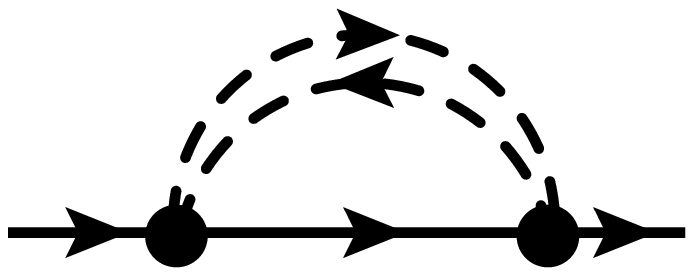}} &
    \includegraphics[width=0.18\textwidth]{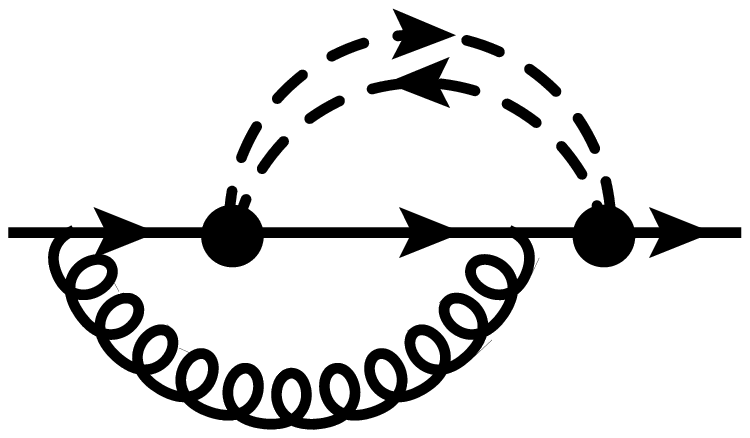} \\
    (a) & (b) \\
    \includegraphics[width=0.18\textwidth]{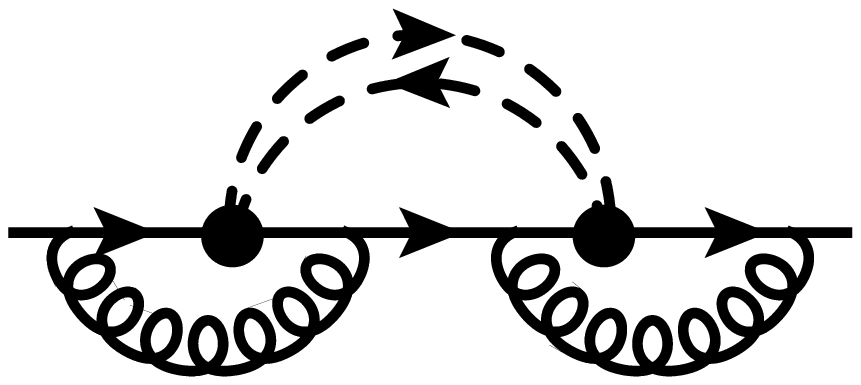} &
    \includegraphics[width=0.18\textwidth]{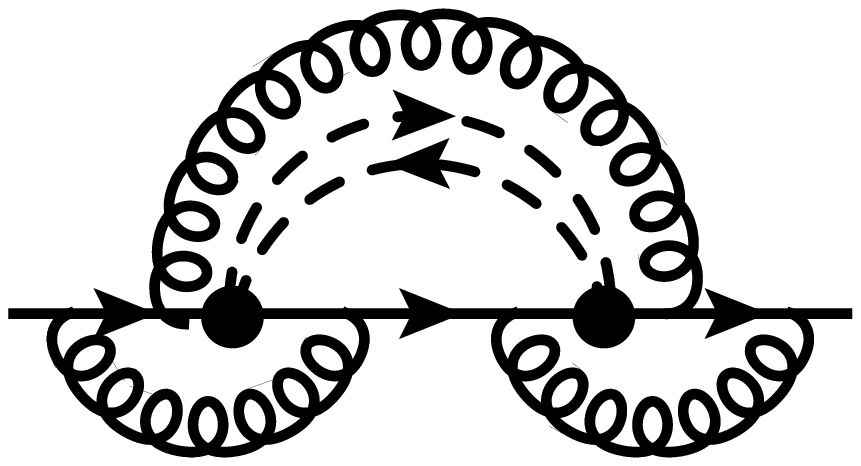} \\
    (c) & (d) \\
    \raisebox{0.3em}{\includegraphics[width=0.18\textwidth]{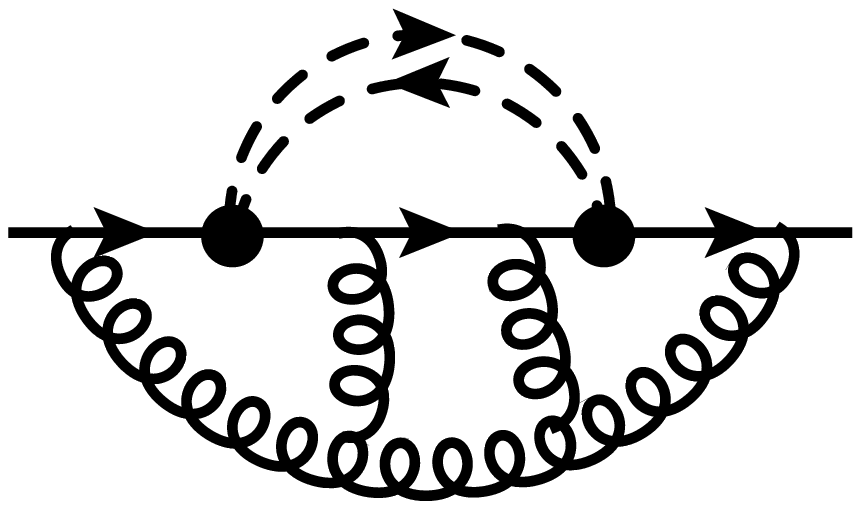}} &
    \includegraphics[width=0.18\textwidth]{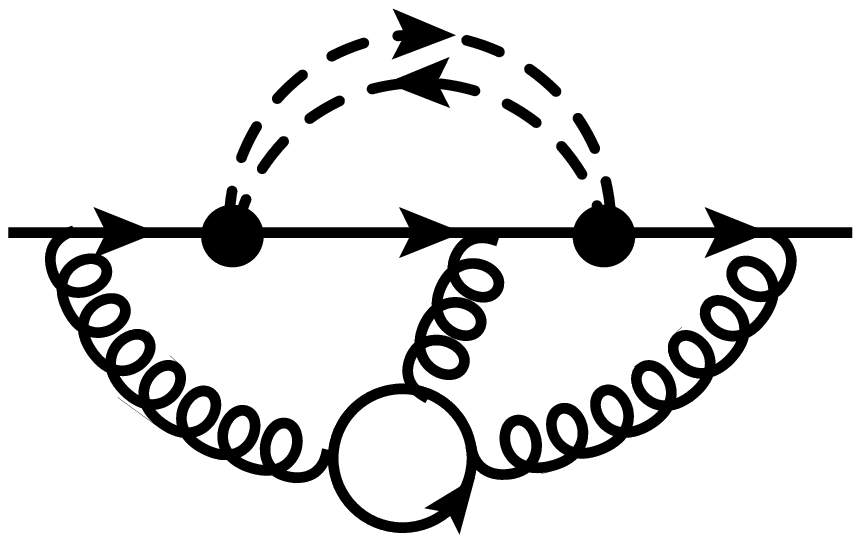} \\
    (e) & (f) \\
  \end{tabular}
  \end{tabular}
  \caption{\label{fig::diag}Sample Feynman diagrams which contribute to the
    forward scattering amplitude of a bottom quark at LO (a), NLO (b), NNLO
    (c) and N$^3$LO (d-f). Straight, curly and dashed lines represent quarks,
    gluons and leptons, respectively. The weak interaction mediated by the $W$
    boson is shown as a blob.}
\end{figure}

 We apply the optical theorem and consider the
forward scattering amplitude of a bottom quark where at leading order the
two-loop diagram in Fig.~\ref{fig::diag}(a) has to be considered. It has a
neutrino, a lepton and a charm quark as internal particles. The weak
interaction is shown as an effective vertex. Our aim is to consider QCD
corrections up to third order which adds up to three more loops. Some sample
Feynman diagrams are shown in Fig.~\ref{fig::diag}(b-f).

The structure of the Feynman diagrams allows the integration of the massless
neutrino-lepton loop which essentially leads to an effective propagator raised
to an $\epsilon$-dependent power, where $d=4-2\epsilon$ is the space-time
dimension. The remaining diagram is at most of
four-loop order.

From the technical point of view there are two basic ingredients which are
crucial to realize our calculation. First, we perform an expansion in the
difference between the bottom and charm quark mass. It has been shown in
Ref.~\cite{Dowling:2008mc} that the expansion converges quite fast for the
physical values of $m_c$ and $m_b$. Second, we apply the so-called method of
regions~\cite{Beneke:1997zp,Smirnov:2012gma} and exploit the similarities to
the calculation of the three-loop corrections to the kinetic
mass~\cite{Fael:2020iea}.

The method of regions~\cite{Beneke:1997zp,Smirnov:2012gma} leads to two
possible scalings for each loop momentum $k^\mu$
\begin{itemize}
\item $|k^\mu| \sim m_b$ ($h$, hard)
\item $|k^\mu| \sim \delta \cdot m_b$ ($u$, ultra-soft)
\end{itemize}
with $\delta=1-m_c/m_b$.  We choose the notion ``ultra-soft'' for the second
scaling to stress the analogy to the calculation of the relation between the
pole and the kinetic mass of a heavy quark,
see~\cite{Fael:2020iea,FSS-mkin-long}.  Note that the momentum which flows
through the neutrino-lepton loop, $\ell$, has to be ultra-soft since
the Feynman diagram has no imaginary part if $\ell$ is hard
since the corresponding on-shell integral has no cut.

Let us next consider the remaining (up to three) momentum integrations which
can be interpreted as a four-point amplitude with forward-scattering
kinematics and two external momenta: $\ell$ and the on-shell momentum
$p^2=m_b^2$. This is in close analogy to the scattering amplitude of a heavy
quark and an external current considered in Ref.~\cite{Fael:2020iea}.
In fact, at each loop order each momentum can either scale as hard or
ultra-soft:
\\[.5em]
\mbox{}\hfill
\begin{tabular}{l|c}
  \hline
  ${\cal O}(\alpha_s)$ & $h, u$ \\
  ${\cal O}(\alpha_s^2)$ & $hh, hu, uu$ \\
  ${\cal O}(\alpha_s^3)$ & $hhh, hhu, huu, uuu$\\
  \hline
\end{tabular}
\hfill\mbox{}
\\[.5em]
Note that all regions where at least one of the loop momenta scales ultra-soft
leads to the same integral families as in
Ref.~\cite{Fael:2020iea,FSS-mkin-long}. The pure-hard regions were absent
in~\cite{Fael:2020iea,FSS-mkin-long}; they lead to (massive) on-shell
integrals.

At this point there is the crucial observation that the integrands in the hard
regions do not depend on the loop momentum $\ell$.  On the other hand, the
ultra-soft integrals still depend on $\ell$. However, for each individual
integral the dependence of the final result on $\ell$ is of the form
\begin{eqnarray}
  (- 2 p \cdot \ell + 2 \delta)^\alpha
\end{eqnarray}
with known exponent $\alpha$.  This means that it is always possible to
perform in a first step the $\ell$ integration which is of the form
\begin{eqnarray}
  \int {\rm d}^d \ell \frac{\ell^{\mu_1}\ell^{\mu_2}\cdots}{(- 2 p \cdot \ell + 2 \delta)^\alpha (-\ell^2)^\beta}
  \,.
  \label{eq::ell-loop}
\end{eqnarray}
A closed formula for such tensor integrals with arbitrary tensor rank and
arbitrary exponents $\alpha$ and $\beta$ can easily be obtained from the
formula provided in Appendix~A of Ref.~\cite{Smirnov:2012gma}.  We thus remain
with the loop integrations given in the above table.  Similar to
Eq.~(\ref{eq::ell-loop}) we can integrate all one-loop hard or ultra-soft
loops which leaves us with pure hard or pure ultra-soft contributions up to
three loops.

A particular challenge of our calculation is the high expansion depth in
$\delta$. We perform an expansion of all diagrams up to $\delta^{12}$.  This
leads to huge intermediate expressions of the order of 
100~GB. Furthermore, for some of the scalar integrals individual
propagators are raised to positive and negative powers up to 12, which
is a non-trivial task for the reduction to master integrals.  For the latter
we combine {\tt FIRE}~\cite{Smirnov:2019qkx} and {\tt
  LiteRed}~\cite{Lee:2012cn}.\footnote{We thank A. Smirnov for providing us
  with the private version of {\tt FIRE} which was crucial for our calculation.}
For the subset of integrals which are needed for the expansion up to
$\delta^{10}$ we also use the stand-alone version of {\tt
  LiteRed}~\cite{Lee:2012cn} as a cross-check.  For all regions where at least
one of the regions is ultra-soft we can take over the master integrals
from~\cite{Fael:2020iea,FSS-mkin-long}.  For some of the (complicated)
three-loop triple-ultra-soft master integrals higher order $\epsilon$ terms
are needed. The method used for their calculation and the results are given
Ref.~\cite{FSS-mkin-long}.  All triple-hard master integrals can be found in
Ref.~\cite{Lee:2010ik}.

%- }}}
%- {{{ Results:

%\bigskip {\bf Results.}
\section{Results}
We write the total decay rate for the $b\to c$ transition in the form
\begin{eqnarray}
  \Gamma(B \to X_c \ell \bar\nu) 
  &=& \Gamma_0 \left[X_0 + C_F\sum_{n\ge 1}
      \left(\frac{\alpha_s}{\pi}\right)^n X_n
      \right] 
      \nonumber\\&&+ {\cal O}\left(\frac{\Lambda_{\rm QCD}^2}{m_b^2}\right)\,,
      \label{eq::gamb2c}
\end{eqnarray}
with $C_F=4/3$, $\Gamma_0=A_{\rm ew}G_F^2|V_{cb}|^2 m_b^5/(192\pi^3)$,
$X_0 = 1 - 8\rho^2 - 12\rho^4 \log(\rho^2) + 8\rho^6 - \rho^8 $ where
$\rho=m_c^{\rm OS}/m_b^{\rm OS}$ and $\alpha_s\equiv\alpha_s^{(5)}(\mu_s)$ with $\mu_s$ being
the renormalization scale. $A_{\rm ew}=1.014$ is the leading electroweak
correction~\cite{Sirlin:1981ie} and $m_b^{\rm OS}$ ($m_c^{\rm OS}$) is the bottom
(charm) pole mass.  The one-
and two-loop results are available from
Refs.~\cite{Trott:2004xc,Aquila:2005hq,Pak:2008qt,Pak:2008cp,Melnikov:2008qs,Biswas:2009rb,Gambino:2011cq,Dowling:2008mc}.
The main result of our calculation is $X_3$. In the following we set all
colour factors to their numerical values.  Furthermore, we specify the number
of massless quarks to 3 and take into account closed charm and bottom
loops. For $\mu=m_b$ we have
\begin{eqnarray}
  X_3 &=& \sum_{n\ge5} x_{3,n} \delta^n\,,
\end{eqnarray}
with analytic coefficients $x_{3,n}$, which in general depend on $\log(\delta)$.
For illustration purposes we show explicit results only for
the leading term which for dimensional reasons is of order $\delta^5$. Our result reads
\begin{align}
  & C_F x_{3,5} =
   \frac{533858}{1215}
  -\frac{20992 a_4}{81}
  +\frac{8744 \pi ^2 \zeta _3}{135}
  -\frac{6176 \zeta_5}{27}
  \nonumber \\ &
  -\frac{16376 \zeta _3}{135}
  -\frac{2624 l_2^4}{243}
  +\frac{5344 \pi ^2 l_2^2}{1215}
  +\frac{179552 \pi ^2l_2}{405}
  \nonumber \\ &
  -\frac{39776 \pi ^4}{6075}
  -\frac{1216402 \pi ^2}{3645}
                 \,,
                 \label{eq::x35}
\end{align}
where $l_2=\log(2)$, $a_4=\mbox{Li}_4(1/2)$ and $\zeta_n$ is the Riemann zeta
function.  Analytic results up to $\delta^{12}$ can be found 
in~\cite{progdata}.  We note that the leading term given in
Eq.~(\ref{eq::x35}) can be cross-checked against the results
from~\cite{Archambault:2004zs} where the $b\to c$ transition has been computed
in the limit $m_c=m_b$.\footnote{After the submission of this paper, the
  authors of Ref.~\cite{Czakon:2021ybq} independently confirmed the terms
  proportional to the $C_F^3$ and $C_F n_h^2$ color factors up to $\delta^9$.}

\begin{figure}[t]
    \includegraphics[width=0.48\textwidth]{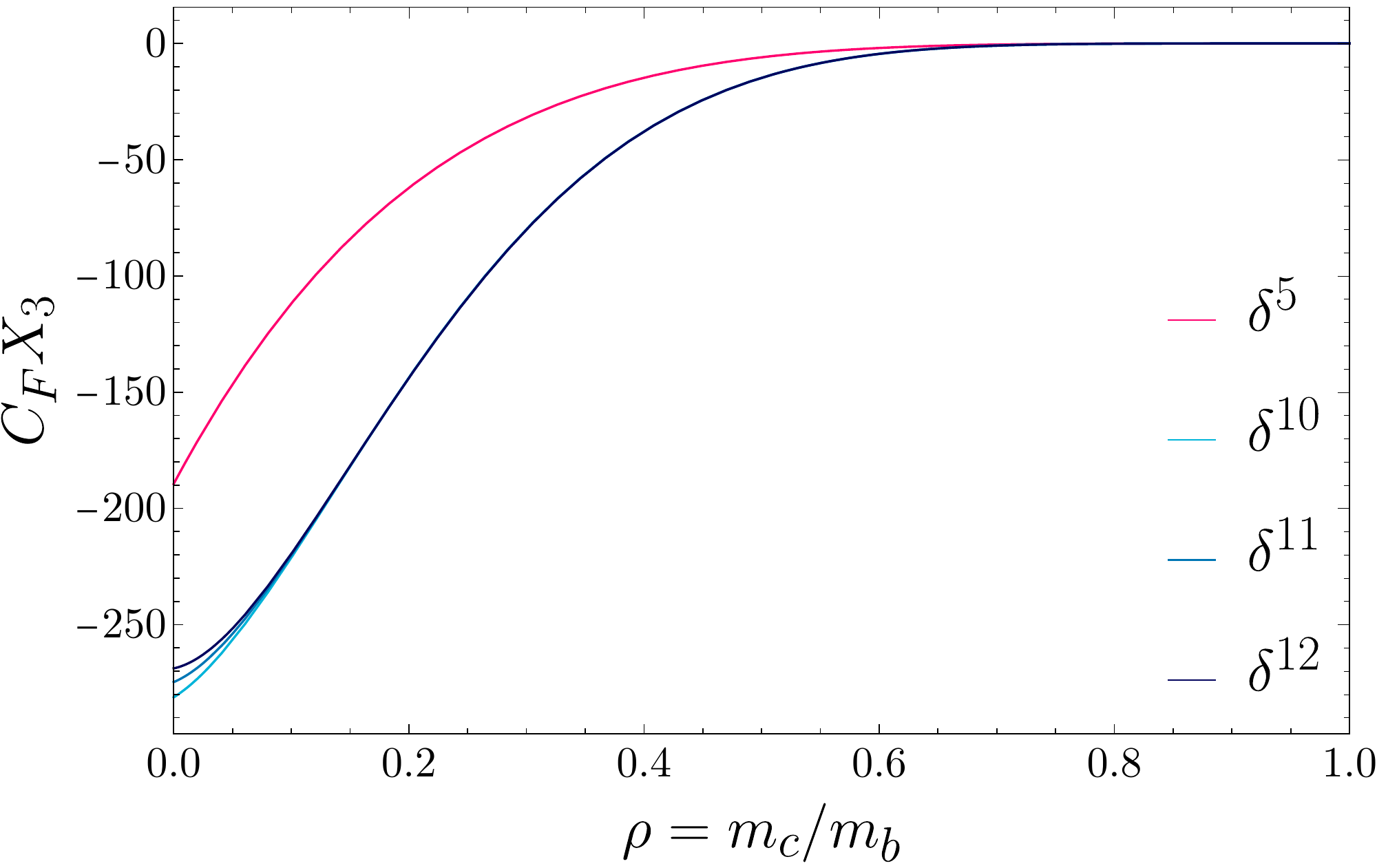}
    \caption{\label{fig::b2c}The third-order coefficient (see
      Eq.~(\ref{eq::gamb2c})) as a function of $\rho=m_c^{\rm OS}/m_b^{
        \rm OS}$ for different
      expansion depth in $\delta$.}
\end{figure}

In Fig.~\ref{fig::b2c} we show $X_3$ as a function of $\rho=1-\delta=m_c^{
  \rm OS}/m_b^{\rm OS}$
where the different curves contain different expansion depths in $\delta$.
One observes a rapid convergence at the physical point for the $b\to c$
decay which amounts to $\rho\approx 0.3$.  In particular, the curves including
terms up to $\delta^{10}$, $\delta^{11}$ or $\delta^{12}$
are basically indistinguishable for $\rho\approx 0.3$ which leads to
$X_3(\rho=0.28)=-{68.4\pm 0.3}$, where the uncertainty is obtained from the
difference of the $\delta^{11}$ and $\delta^{12}$ expansion, multiplied
by a security factor of five.

\begin{table}[t]
  \begin{tabular}{c|rrrrr}
    \hline
    & $Y_1$& $Y_2^{\rm rem}$& $\beta_0 Y_2^{\beta_0}$& $Y_3^{\rm rem}$& $\beta_0^2 Y_3^{\beta_0^2}$ \\
    \hline
    $m_b^{\rm OS},m_c^{\rm OS}$
    & $-1.72$ &$3.08$ &$-16.17$ &$48.8$ &$-212.1$ \\
    $m_b^{\rm kin}, m_c^{\rm kin}$
    & $-0.94$ &$0.33$ &$-4.08$ &$-5.4$ &$-15.4$ \\
    $m_b^{\rm kin}, \overline{m}_c(3~\mbox{GeV})$
    & $-1.67$ &$-3.39$ &$-3.85$ &$-97.7$ &$69.1$ \\
    $m_b^{\rm kin}, \overline{m}_c(2~\mbox{GeV})$
    & $-1.25$ &$-1.21$ &$-2.43$ &$-68.8$ &$67.9$ \\
    $\overline{m}_b(\overline{m}_b), \overline{m}_c(3~\mbox{GeV})$
    & $3.07$ &$-21.81$ &$35.17$ &$-56.7$ &$119.4$ \\
    $m_b^{\rm PS}, \overline{m}_c(2~\mbox{GeV})$
    & $- 0.47$ &$-6.10$ &$- 2.31$ &$-93.1$ &$- 7.19$ \\
    $m_b^{\rm 1S}, \overline{m}_c(m_b^{\rm 1S})$
    & $- 3.59$ &$-0.98$ &$- 19.39$ &$-39.83$ &$- 80.22$ \\
    $m_b^{\rm 1S}, m_c$ via HQET
    & $- 1.38$ &$0.73$ &$-7.05$ &$5.04$ &$-38.09$ \\
    \hline
  \end{tabular}
  \caption{\label{tab::Yn}Numerical results for the coefficients $Y_n$ in
  Eq.~(\ref{eq::gamb2c-2}) for various renormalization schemes.}
\end{table}

For the numerical evaluation it is convenient to cast Eq.~(\ref{eq::gamb2c})
in the form
\begin{eqnarray}
  \Gamma(B \to X_c \ell \bar\nu) 
  &=& \Gamma_0 X_0 \left[ 1 + \sum_{n\ge 1}
      \left(\frac{\alpha_s}{\pi}\right)^n Y_n\right]
      \nonumber\\&&\mbox{}
      + {\cal O}\left(\frac{\Lambda_{\rm QCD}^2}{m_b^2}\right)\,,
      \label{eq::gamb2c-2}
\end{eqnarray}
with $\alpha_s\equiv \alpha_s^{(4)}(\mu_s)$ as expansion parameter.  In the
following we discuss various renormalization schemes for the charm and bottom
quark masses, where $\Gamma_0$ and $X_0$ are evaluated using the respective
numerical values.  In Tab.~\ref{tab::Yn} we provide the corresponding
results for the coefficients $Y_n$. At two and three-loop orders we split the
results into the large-$\beta_0$ contribution and the remaining term
\begin{eqnarray}
  Y_2 &=& Y_2^{\rm rem} + \beta_0 Y_2^{\beta_0} \,,\nonumber\\
  Y_3 &=& Y_3^{\rm rem} + \beta_0^2 Y_3^{\beta_0^2} \,,
\end{eqnarray}
with $\beta_0 = 11 - 2/3n_l = 9$ where $n_l=3$ is the number of massless
quarks.  Note that the uncertainty of
$Y_3$ due to the expansion in $\delta$ is of the same order of magnitude as
for $X_3$ discussed above.

For the transition of the on-shell quark masses to the $\overline{\rm MS}$
scheme we use the three-loop formulae provided in
Refs.~\cite{Chetyrkin:1999qi,Melnikov:2000qh}.  Finite-$m_c$ effects in the
bottom mass relation are taken from Refs.~\cite{Fael:2020bgs}. The two- and
three-loop corrections to the transition from the on-shell to the kinetic
scheme are provided in~\cite{Czarnecki:1997sz}
and~\cite{Fael:2020iea,FSS-mkin-long}, respectively. Note that the transition
to the kinetic scheme also requires the renormalization of the parameters
$\mu_\pi^2$ and and $\rho_D^3$, which enter the decay rate at order $1/m_b^2$
and $1/m_b^3$, respectively.  They receive additive contributions, which enter
$Y_i$ in Eq.~(\ref{eq::gamb2c-2})~\cite{Benson:2003kp,Gambino:2007rp}. The
corresponding corrections up to three-loop order can be found
in~\cite{FSS-mkin-long}.  Note that we assume a heavy charm quark and thus we
have $(n_l=3)$-flavour QCD as starting point for the on-shell--kinetic
relations.  We use the decoupling relation for $\alpha_s$ up to two-loop order
to obtain expressions parameterized in terms of $\alpha_s^{(4)}$.  For the
decoupling scale we use $\mu_s$. It has been shown in
Ref.~\cite{FSS-mkin-long} that there are no additional charm quark mass
effects in the kinetic-on-shell relation.  For comparison we show in
Tab.~\ref{tab::Yn} also results where the bottom quark mass is renormalized in
the PS~\cite{Beneke:1998rk} and
1S~\cite{Hoang:1998hm,Hoang:1998ng,Hoang:1999zc} scheme.  In the latter case
we renormalize the charm quark mass both in the $\overline{\rm MS}$ and via
the Heavy Quark Effective Theory (HQET) relation to on-shell bottom quark mass
and (averaged) $D$ and $B$ meson masses (see, e.g., Ref.~\cite{Hoang:1998ng}).
After each scheme change we re-expand in $\alpha_s$ to third order.

Note that our two-loop results for
$Y_2^{\rm rem}$ differ from the one of Ref.~\cite{Alberti:2014yda} due to
finite charm quark mass effects in the relation between the kinetic and
on-shell bottom quark mass and the renormalization of $\mu_\pi^2$ and
$\rho_D^3$~\cite{FSS-mkin-long}. This leads to a shift of about 
$-0.5$\% in the leading $1/m_b$ approximation of the decay rate and thus
might have a visible effect on the value of $|V_{cb}|$.

For the numerical evaluation of the decay rate we use
the input values $m_b^{\rm OS}=4.7$~GeV,
$m_c^{\rm OS}=1.3$~GeV, $m_b^{\rm kin}=4.526$~GeV,
$m_b^{\rm PS}=4.479$~GeV, $m_b^{\rm 1S}=4.666$~GeV,
$m_c^{\rm kin}=1.130$~GeV,
$\overline{m}_b(\overline{m}_b)=4.163$~GeV,
$\overline{m}_c(3~\mbox{GeV}) =0.993$~GeV,
$\overline{m}_c(2~\mbox{GeV}) =1.099$~GeV, and
$\alpha_s^{(5)}(M_Z)=0.1179$. We use {\tt RunDec}~\cite{Herren:2017osy} for
the running of the $\overline{\rm MS}$ parameters and the decoupling of heavy
particles.  For the Wilsonian cutoff in the kinetic scheme we use $\mu=1$~GeV
both for the bottom and charm quark. In the case of PS scheme
we use $\mu=2$~GeV. For the renormalization scale of
$\alpha_s^{(4)}$, $\mu_s$, we choose the respective value for the bottom quark
mass.

For illustration purpose we provide in Tab.~\ref{tab::Yn} also results where
both masses are defined in the on-shell scheme. It is well known that in this
scheme the perturbative series shows a bad convergence behaviour.  In fact, we
have $Y_3\approx -163$ whereas in the schemes where the bottom quark
mass is used in the kinetic scheme we have that $Y_3$ is between $-1$ and
$-29$.  Note, that in the scheme where both quark masses are defined in
the $\overline{\rm MS}$ scheme the three-loop corrections are more than twice
as big which also hints for a worse convergence behaviour. 
The PS and 1S schemes show a clear improvement as compared to the on-shell
scheme. However, the convergence properties are significant worse than in the
kinectic scheme in case the charm quark mass is renormalized in the
$\overline{\rm MS}$ scheme. In case $m_c^{\rm OS}$ is expressed through
$m_b^{\rm OS}$ and meson masses using a HQET
relation one observes an improved perturbative behaviour.
Still, the analysis
clearly shows the advantage of the kinetic scheme which is constructed such
that large corrections are resummed into the quark mass value.  In fact, all
three schemes which involve $m_b^{\rm kin}$ demonstrate a good convergence
behaviour. Using $\alpha_s^{(4)}(m_b^{\rm kin})=0.2186$ we obtain for
$\Gamma(B \to X_c \ell \bar\nu) / \Gamma_0$ in these three schemes
\begin{eqnarray}
  m_b^{\rm kin}, m_c^{\rm kin}: &\hspace*{0.5em}& 
  0.633 \left( 
    1 -0.066-0.018-0.007 \right)
\nonumber\\&&\approx 0.575 \,, \nonumber\\
  m_b^{\rm kin}, \overline{m}_c(3~\mbox{GeV}): &&
  0.700 \left( 
    1 -0.116-0.035-0.010 \right)
\nonumber\\&&\approx 0.587 \,, \nonumber\\
  m_b^{\rm kin}, \overline{m}_c(2~\mbox{GeV}): &&
  0.648 \left( 
    1 -0.087-0.018-0.0003 \right)
\nonumber\\&&\approx 0.580 \,, 
\end{eqnarray}
where the different $\alpha_s$ orders are displayed
separately. Note that in the PS and 1S schemes the third-order corrections
amount to 3.4\% and 3.9\%, respectively, with $m_c$ in the $\overline{\rm MS}$
scheme.
If one defines $m_c$ in the 1S scheme via a HQET relation the third-order corrections
reduce to 1\%. For the bottom mass expressed in the kinetic scheme
we observe that the third-order corrections amount to at most 1\%
and they are a factor two to three smaller than the corrections of order
$\alpha_s^2$. A particularly good behaviour is observed for the choice
$\overline{m}_c(2~\mbox{GeV})$ where the corrections of order $\alpha_s^3$
are below the per mille level. Its final result lies
between the other two kinetic schemes and deviates from them by about 0.9\%
and 1.2\%, respectively.

\begin{figure}[t]
  \begin{tabular}{c}
  \includegraphics[width=0.48\textwidth]{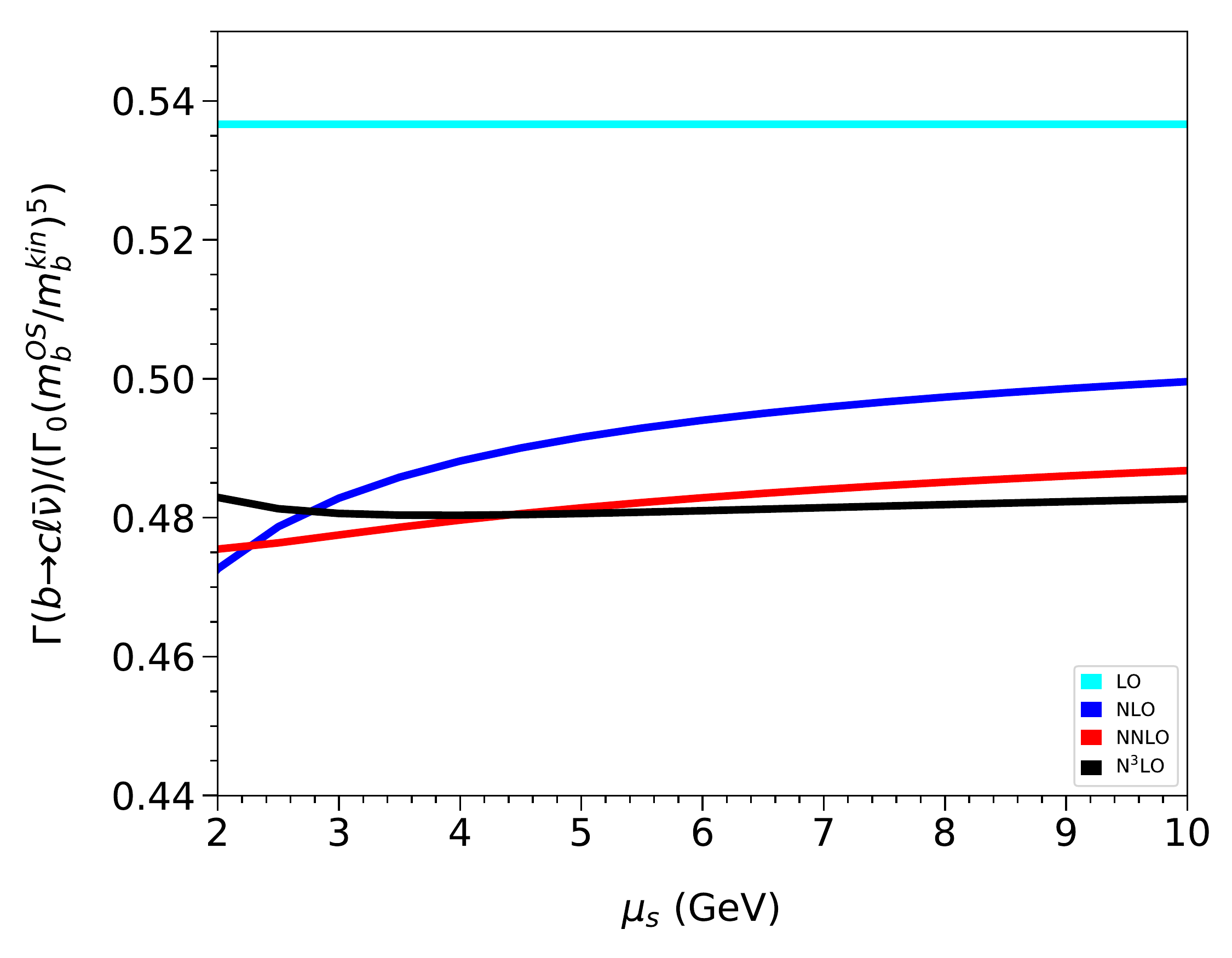}
  \\
    (a)
  \\
  \includegraphics[width=0.48\textwidth]{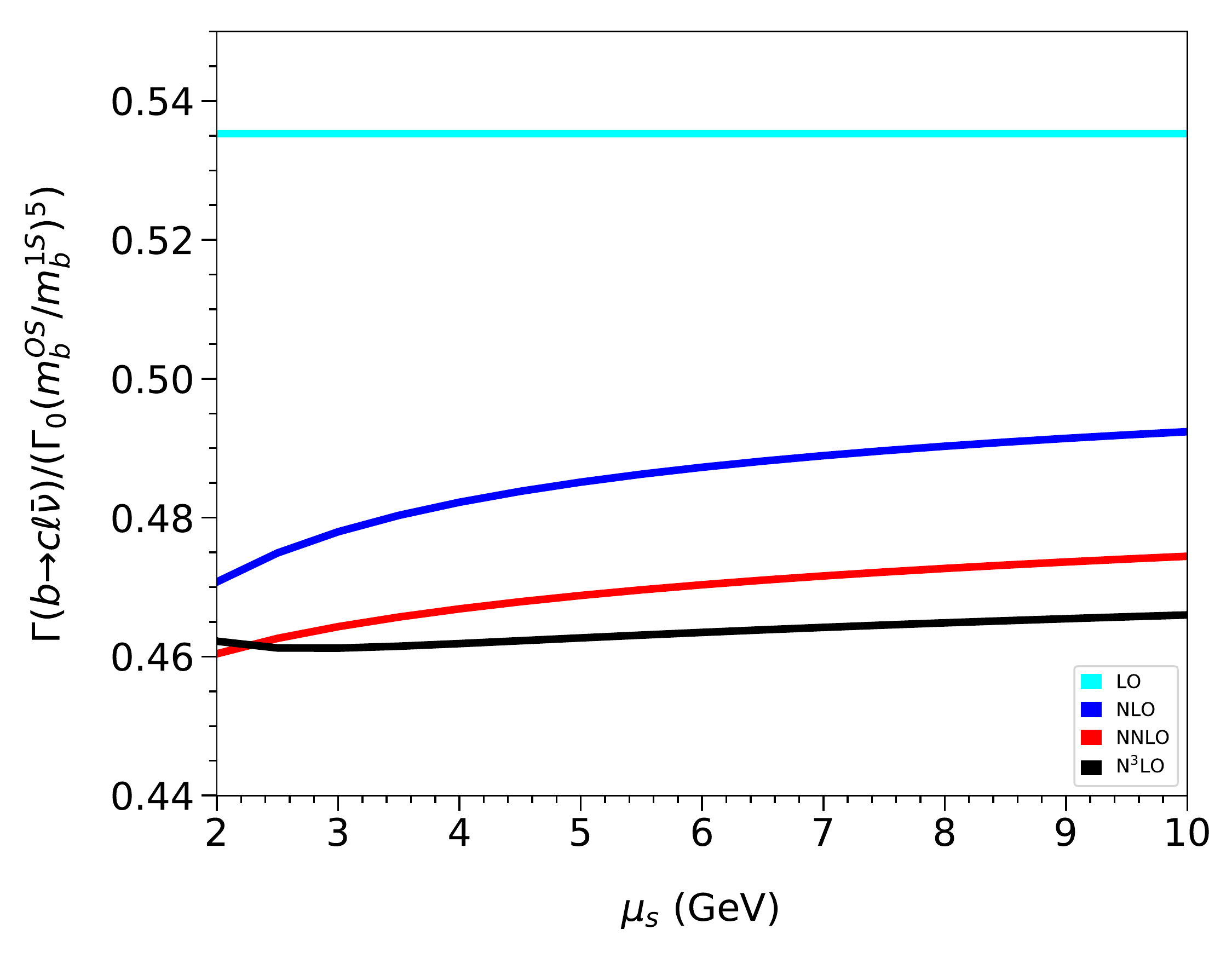}
    \\
    (b)
  \end{tabular}
  \caption{\label{fig::gam_mus}Total partonic decay rate in the 
    kinetic (a) and 1S scheme (b) as a function of the renormalization scale $\mu_s$.
    See text for details.
    Note that the normalization chosen for the $y$ axis is scheme independent.}
\end{figure}

In Fig.~\ref{fig::gam_mus} we show the partonic decay rate as a function of
the renormalization scale $\mu_s$. Fig~\ref{fig::gam_mus}(a) shows the bottom
quark mass renormalized in the kinetic and the charm quark mass in the
$\overline{\rm MS}$ scheme.  One observes that over the whole range
$2~\mbox{GeV}<\mu_s<10~\mbox{GeV}$ the dependence on $\mu_s$ decreases after
including higher order corrections.  (The LO order result is
$\mu_s$-independent by construction.) Whereas at NNLO one observes still a
2.5\% variation, it is far below the percent level at N$^3$LO.
Fig~\ref{fig::gam_mus}(b) shows the corresponding results for the 1S scheme
where $m_c$ is defined via a HQET relation.

The total partonic rate in the kinetic and in the 1S scheme differ for the
following reason. Higher power corrections are not included in our partonic
$b\to c \ell \bar \nu_\ell$ prediction. In particular the kinetic scheme
absorbs $\mu^2/m_b^2$ and $\mu^3/m_b^3$ terms from the redefinition of
$\mu_\pi^2$ and $\rho_D^3$, while in the 1S scheme we neglect higher $1/m_b$
and $1/m_c$ power corrections when expressing the charm mass in terms of meson
masses within HQET. Only the $B\to X_c \ell \bar \nu_\ell$ total rate
predictions can be compared.

In general the large-$\beta_0$ terms provide dominant contributions.
However, in all cases the remaining terms are not negligible and often 
have a different sign. In the kinetic
scheme where the charm quark is renormalized in the $\overline{\rm MS}$ scheme
the remaining contributions are numerically even bigger than the 
large-$\beta_0$ terms.

It is impressive that the expansion in $\delta$ shows a good converge
behaviour even for $\delta\to1$ which corresponds to a massless daughter
quark. This allows us to extract the coefficient $X_3$ for the decay
$b\to u\ell\bar\nu$.  A closer look to the $\delta^{10}$, $\delta^{11}$, and
$\delta^{12}$ terms in Fig.~\ref{fig::b2c} indicates that the convergence is
quite slow for $\rho\to0$.  As central value for the three-loop prediction we
use our approximation based on the $\delta^{12}$ term and estimate the
uncertainty from the behaviour of the one- and
two-loop~\cite{vanRitbergen:1999gs,Steinhauser:1999bx} results for $\rho=0$,
where the exact results are known.  Incorporating expansion terms up to order
$\delta^{12}$ we observe a deviation of about 3.5\% whereas the $\delta^{12}$
terms amount to less than 1\%, both at one and two loops. At three loops the
$\delta^{12}$ term amounts to about 2\%. We thus conservatively estimate the
uncertainty to 10\% which leads to
\begin{eqnarray}
  X_3^u &\approx& -202 \pm 20\,.
\end{eqnarray}
In this result the contributions with closed charm loops are approximated with
$m_c=0$.

\begin{figure}[t]
  \includegraphics[width=0.48\textwidth]{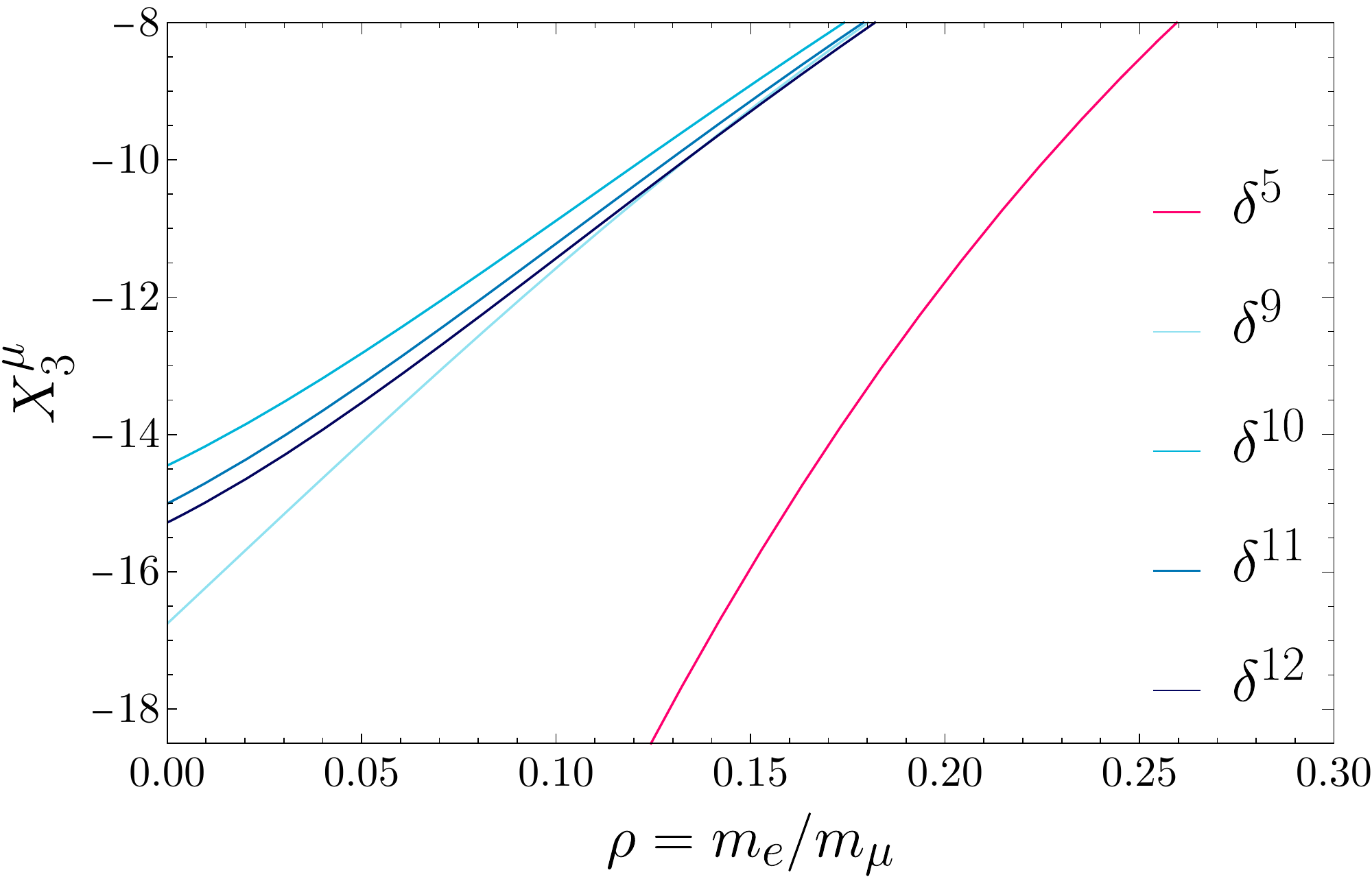}
  \caption{\label{fig::m2e}The third-order coefficient to $\Delta q$
    introduced in Eq.~(\ref{eq::tau_mu}) as a function of $m_e/m_\mu$.}
\end{figure}

In the remaining part of this paper we specify our results to QED and study
the corrections to the muon decay.  A comprehensive review of the various
correction terms is given in Ref.~\cite{vanRitbergen:1999fi} where $\Delta q$
in Eq.~(\ref{eq::tau_mu}) is parameterized as
\begin{eqnarray}
  \Delta q &=& \sum_{i\ge 0} \Delta q^{(i)}\,.
\end{eqnarray}
$\Delta q^{(0)}$ is given by $X_0-1$ (see Eq.~(\ref{eq::gamb2c})) with
$\rho = m_e/m_\mu$ and $\Delta q^{(1)}$~\cite{Kinoshita:1958ru} and
$\Delta q^{(2)}$~\cite{vanRitbergen:1998yd,Steinhauser:1999bx} are easily
obtained after specification of the QCD colour factors to their QED values
(see Ref.~\cite{vanRitbergen:1999fi} for analytic results). We introduce
$\Delta q^{(3)} = (\alpha(m_\mu)/\pi)^3 X_3^\mu$, where $\alpha(m_\mu)$ is
the fine structure constant in the $\overline{\rm MS}$
scheme~\cite{vanRitbergen:1999fi}. In Fig.~\ref{fig::m2e} 
we show the third-order coefficient $X_3^\mu$ for $0\le \rho\le 0.3$.  At the
physical point $m_e/m_\mu\approx 0.005$ the convergence behaviour is similar
to QCD.  We estimate $X_3^\mu$ using the same approach as for $X_3^u$
and examine the one- and two-loop behaviour. Up to an overall factor $C_F$ the
one-loop term is, of course, identical to the $b\to u$ transition.  Including
expansion terms up to $\delta^{12}$ at two loops leads to a deviation by about
8\% from the exact result whereas the $\delta^{12}$ term itself contributes by
about 1\%. The three-loop $\delta^{12}$ amounts to about 2\%. Assuming the
same relative contribution thus leads to an uncertainty estimate of about 15\%
and we have
\begin{eqnarray}
  \Delta q^{(3)} 
  &\approx& 
            \left(\frac{\alpha(m_\mu)}{\pi}\right)^3 \left( - 15.3 \pm 2.3 \right)
            \,.
\end{eqnarray}
In Ref.~\cite{Ferroglia:1999tg} the three-loop corrections were
estimated to $X^\mu_3 \sim -20$.  With the help of
Eq.~(\ref{eq::tau_mu}) we obtain for the $\alpha^3$ QED contribution
to the muon life time ${(-9 \pm 1) \times 10^{-8}}$~$\mu$s.  This
result has to be compared to the current experimental value which is
given by $\tau_\mu = 2.1969811 \pm
0.0000022$~\mbox{$\mu$s}~\cite{Tanabashi:2018oca}.  The new correction
terms are almost two orders of magnitude smaller than the 
experimental uncertainty. Thus, an  updated value of $G_F$ can only be
extracted once the latter has been improved.

%- }}}
%- {{{ Concl.:

%\bigskip {\bf Conclusions.}
\section{Conclusions}
We have computed
three-loop corrections of order $\alpha_s^3$ to the total decay rate
$\Gamma(B\to X_c\ell \bar{\nu})$ including finite charm quark mass effects. We
perform an expansion around the equal-mass case and demonstrate that a
good convergence at the physical point is observed after taking into
account eight expansion terms.  Our result is one of the very few
third-order results to physical quantities available to date involving
two different mass scales.

We can extend our considerations to the case of a massless charm quark
and thus obtain corrections of order $\alpha_s^3$ to $\Gamma(B\to
X_u\ell \bar{\nu})$, although with a larger uncertainty of about $10\%$. After
specifying our findings to QED we furthermore obtain predictions for
the third-order corrections to the muon decay. Here we estimate the
uncertainty to 15\%.

The decay rate $\Gamma(B\to X_c\ell \bar{\nu})$ is an important ingredient for
the determination of the CKM matrix element $|V_{cb}|$. However, a
detailed analysis (see, e.g., Ref.~\cite{Alberti:2014yda}) also
requires the knowledge of moments of kinematic distributions.  The
method described in this paper can also be applied to the calculation
of such moments at order $\alpha_s^3$, although at the cost of
significantly increased computer resources.

%- }}}

\smallskip

{\bf Acknowledgements.}  We thank Paolo Gambino for communications and
clarifications concerning Ref.~\cite{Alberti:2014yda}. We are grateful
to Alexander Smirnov for his support in the use of {\tt FIRE} and to
Florian Herren for providing us his program {\tt
  LIMIT}~\cite{Herren:2020ccq} which automates the partial fraction
decomposition in case of linearly dependent denominators. We thanks
Joshua Davies for valuable advice in optimizing the usage of {\tt
  FORM}~\cite{Ruijl:2017dtg}. This research was supported by the Deutsche
Forschungsgemeinschaft (DFG, German Research Foundation) under grant
396021762 --- TRR 257 ``Particle Physics Phenomenology after the Higgs
Discovery''.

\end{document}